\documentstyle[twoside,fleqn,espcrc2,epsf]{article}

\newcommand{\AmS}{{\protect\the\textfont2
  A\kern-.1667em\lower.5ex\hbox{M}\kern-.125emS}}

\title{Topics in Light Hadron Mass Spectrum in Quenched QCD
\thanks{Presented by T. Yoshi\'e}}

\author{JLQCD Collaboration \\[2mm]
        S.~Aoki\address{Institute of Physics, University of
        Tsukuba, Tsukuba, Ibaraki 305, Japan},
        M.~Fukugita\address{Yukawa Institute for Theoretical Physics,
        Kyoto University, Kyoto 606, Japan},
        S.~Hashimoto\address{National Laboratory for High
        Energy Physics (KEK), Tsukuba, Ibaraki 305, Japan},
        N.~Ishizuka$^{\rm a}$,
        Y.~Iwasaki$^{\rm a,}$\address{Center for Computational Physics,
        University of Tsukuba, Tsukuba, Ibaraki 305, Japan},
        K.~Kanaya$^{\rm a,d}$,
        Y.~Kuramashi$^{\rm c}$,
        H.~Mino\address{Faculty of Engineering, Yamanashi
        University, Kofu 400, Japan},
        M.~Okawa$^{\rm c}$, A.~Ukawa$^{\rm a}$,
        T.~Yoshi\'e$^{\rm a,d}$ }

\begin{document}

\begin{abstract}
Several topics concerning the light hadron spectrum are discussed.
Flavor symmetry breaking effects and the problem of quenched chiral logarithm
are examined with pion mass data for the Kogut-Susskind quark action, and   
light meson decay constants for the Wilson action calculated with 
non-perturbative renormalization constants are discussed.  
Results for quark masses are also given
both for the Kogut-Susskind and Wilson actions. 
\end{abstract}

\maketitle

\section{Introduction}

In this report we discuss several issues 
concerning the light hadron mass spectrum 
in quenched QCD with the Kogut-Susskind~(KS) and Wilson quark actions. 
For the KS action we use data generated in our $B_K$ measurements
in the range $\beta=5.7-6.4$~\cite{aoki}.
 The Wilson data come from $f_B$~\cite{hashimoto} and
 $B_K$~\cite{kuramashi} measurements at $\beta=5.9-6.3$.
We refer to refs.~\cite{aoki,hashimoto,kuramashi} 
for simulation details, concentrating here on physics issues.
  
\section{Flavor breaking with the KS action}

\begin{figure}[t]
\begin{center}
\leavevmode
\epsfxsize=7.5cm 
  \epsfbox{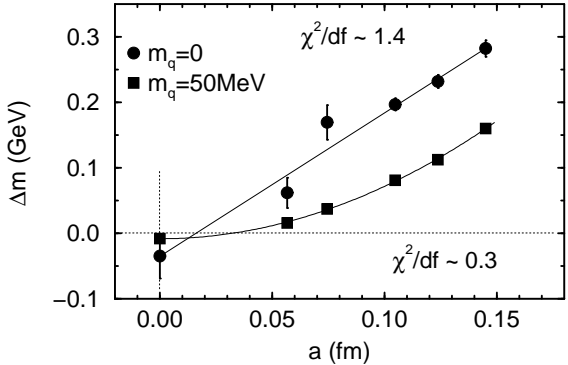}
\end{center}
\vspace{-1.4cm}
\caption{$\Delta m_\pi = m_{\pi(\gamma_5\gamma_4\otimes\xi_5\xi_4)} -
m_{\pi(\gamma_5\otimes\xi_5)}$ as a function of $a$. Lines are 
linear ($m_q=0$) or quadratic ($m_q$= 50 MeV) fits. }
\label{fig1}
\vspace{-10mm}
\end{figure}

Spectral quantities for the KS action are expected
to have flavor breaking errors of $O(a^2)$.
For the pion spectrum in physical units this means that
\begin{eqnarray}
&&m_{\pi(\mbox{\small NG})}^2=Bm_q+O(m_q^2)\nonumber\\
&&m_{\pi(\mbox{\small non-NG})}^2=\Lambda^4 a^2+Bm_q+O(m_q^2)\nonumber
\end{eqnarray}
to leading order in $a$, where $\Lambda$ denotes the symmetry-breaking scale, 
{\it NG} means the Nambu-Goldstone channel 
$\gamma_5\otimes \xi_5$ and {\it non-NG} any other channels. 
An interesting implication is that the mass splitting 
$\Delta m_\pi=m_{\pi(\mbox{NG})}-m_{\pi(\mbox{non-NG})}$ 
should vanish linearly $\Delta m_\pi=O(a)$ at $m_q=0$, 
while the behavior should change to $\Delta m_\pi=O(a^2)$ 
for fixed and finite values of $m_q$ satisfying $Bm_q \gg \Lambda^4a^2$. 
These predictions are confirmed in our data 
as shown in fig.~\ref{fig1}, 
where we take $\pi(\gamma_5\gamma_4\otimes\xi_5\xi_4)$ 
for the non-Nambu-Goldstone pion.  
This conforms the result of a similar test made previously~\cite{sharpe91}.

\section{Quenched chiral logarithm}

The Nambu-Goldstone pion mass for the KS action offers 
a good testing ground for the predictions of quenched chiral 
perturbation theory~\cite{SBGXLOG}.
We fit our data for $m_{\pi(\gamma_5\otimes\xi_5)}$ 
for equal and unequal quark masses to the form (fit-1),
%
\begin{eqnarray*}
(m_\pi a)^2 &=& A(m_1+m_2)a \biggr[ 1 
  - \delta \cdot \biggr\{ \log {2A m_1\over \Lambda} \\
  &+& {m_2\over m_2-m_1}\log{m_2\over m_1}\biggl\} \biggr] \\
  &+& C_4 (m_1+m_2)^2a^2 + C_5 m_1m_2a^2. \\
\end{eqnarray*}
The quadratic terms are added since fits with only the logarithm
term become untenable as $\beta$ increases beyond $\beta=6.0$. 
Fits are also made in which the quark masses inside the logarithm 
are corrected by a constant shift to take into account 
the non-Nambu-Goldstone nature of $\eta^\prime$ in the KS formalism
(fit-2).

\begin{figure}[t]
\begin{center}
\leavevmode
\epsfxsize=7.5cm 
  \epsfbox{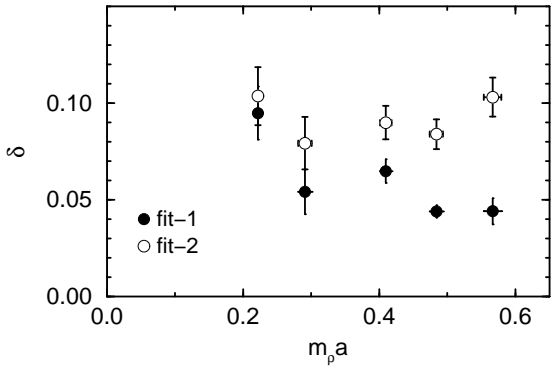}
\end{center}
\vspace{-1.4cm}
\caption{Parameter $\delta$ extracted from the Nambu-Goldstone pion as a
function of $m_\rho a$.
See text for details.}
\label{fig2}
\vspace{-10mm}
\end{figure}

We find both types of fits to be acceptable, and the resulting $\delta$
reasonably stable as a function of lattice spacing as shown in 
fig.~\ref{fig2}.  In this sense our data are consistent with the presence of 
the quenched chiral logarithm.  
The magnitude of $\delta$, 
$\delta=0.05-0.07$ for fit-1 and  
$\delta=0.08-0.1$ for fit-2, 
however, is a factor 2--3 smaller 
than is expected in the real world $\delta\approx 0.2$.

\section{Light Quark Masses}

\begin{figure}[t]
\begin{center}
\leavevmode
\epsfxsize=7.5cm 
  \epsfbox{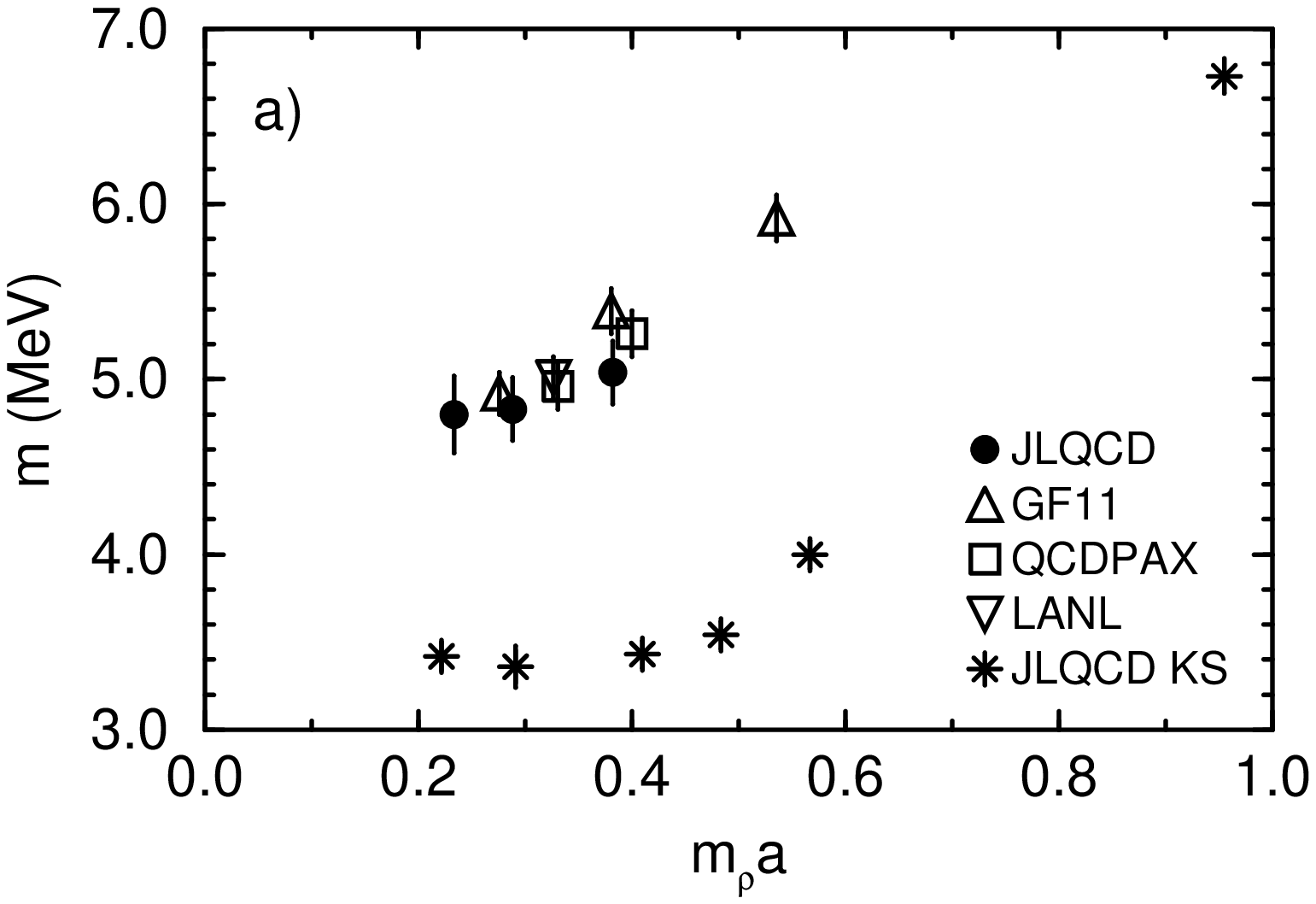}
\end{center}
\begin{center}
\leavevmode
\epsfxsize=7.5cm 
  \epsfbox{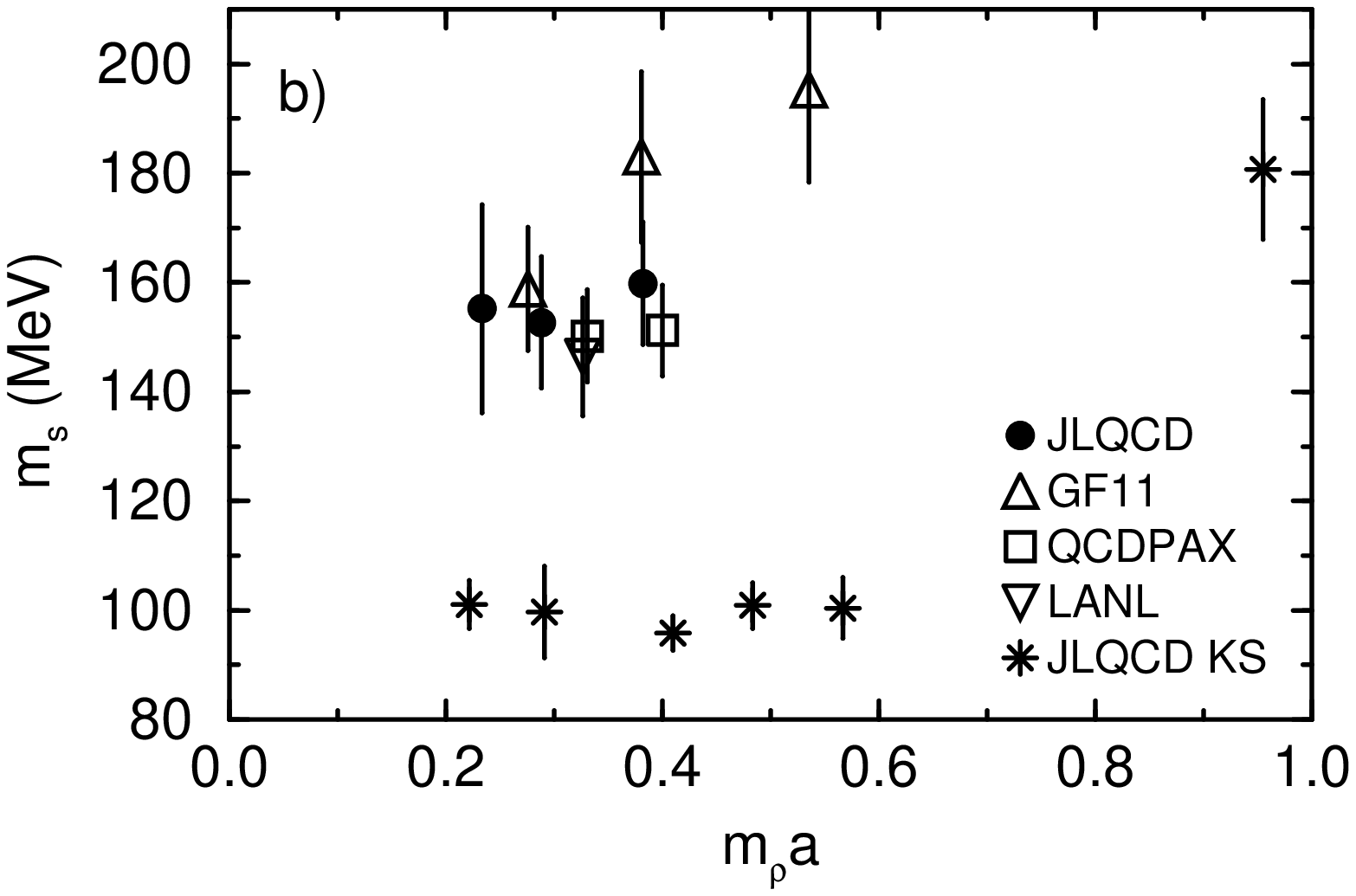}
\end{center}
\vspace{-1.4cm}
\caption{(a) Average of up and down quark masses 
in the $\overline{MS}$ scheme at 2 GeV. 
Representative Wilson results from other
groups~\protect\cite{GF11,QCDPAX96,LANL96} 
recalculated with the same procedure are also shown. 
(b) Strange quark mass estimated from $m_\phi$.} 
\label{fig3}
\vspace{-8mm}
\end{figure}

Continuum extrapolations of light quark masses have been recently discussed by 
several authors~\cite{mackenzie,LANLMQ}.
In fig.~\ref{fig3}(a) we plot our results for $\overline
m = (m_u + m_d)/2$ in the $\overline {MS}$ scheme 
at $\mu =$ 2 GeV for the Wilson
(solid circles)~\cite{hashimoto} and KS (bursts)~\cite{aoki} actions.
Results for the strange quark mass
determined from $m_\phi$ are shown in fig.~\ref{fig3}(b) 
(estimates from $m_K$ are not shown, 
being essentially $m_s\approx 26\overline{m}$).  Our results are
obtained by a 2-loop running of tadpole-improved $\overline{MS}$ masses at the
scale $1/a$ calculated with $\alpha_V(1/a)$.  The lattice spacing is fixed 
by $m_\rho$.  

For the Wilson action our results show a weaker $a$ dependence 
compared with those of other groups, albeit errors are large and the lattice 
size used is rather small ($La\approx 1.9$fm). 
As a result, the agreement of Wilson and KS results 
after continuum extrapolation is less apparent ({\it cf.} ref.~\cite{LANLMQ}).
We note, however, that the one-loop renormalization factor 
for the KS action has a large value of almost two in the range of our data.
Higher order corrections might well bring the results for the KS action
into agreement with those of the Wilson action.  

\section{Meson decay constants for the Wilson action}

Determination of hadronic matrix elements often suffers from uncertainties in
renormalization factors.  An exceptional case is the vector meson decay
constant $f_V^{-1}$ for which the renormalization constant 
$Z_V$ for the local vector current can be precisely determined 
non-perturbatively with the use of the conserved current~\cite{MMZMC}.
For our Wilson runs at $\beta =$ 5.9, 6.1 and 6.3~\cite{hashimoto},
we obtain $Z_V$= 0.532(4), 0.595(3) and 0.640(4), respectively.

\begin{figure}[t]
\begin{center}
\leavevmode
\epsfxsize=7.5cm 
  \epsfbox{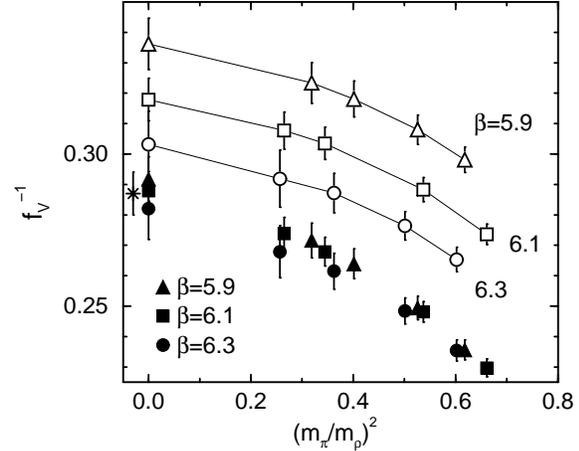}
\end{center}
\vspace{-1.4cm}
\caption{Vector meson decay constants for the Wilson quarks.
Open (filled) symbols are those with tadpole improved
(non-perturbatively determined) $Z$ factors.}
\label{fig4}
\vspace{-7mm}
\end{figure}

In fig.~\ref{fig4} we plot $f_V^{-1}$ as a function of $(m_\pi/m_\rho)^2$ for
the three $\beta$ values using the non-perturbative $Z_V$
as well as the tadpole-improved perturbative result 
$Z_V = (1-0.82 \alpha_V(1/a))/(8K_c)$.
We find an almost perfect scaling with the non-perturbative estimate
of $Z_V$ as was noted previously~\cite{QCDPAX96}, 
while the perturbative $Z$ factor gives results 
significantly varying with $\beta$. 

Making a linear extrapolation of the non-perturbative results in $m_\rho a$, 
we obtain
$f_\rho^{-1} =$ 0.270(25) and $f_\phi^{-1} =$ 0.239(10)
in the continuum limit.  These values are in a good agreement
with the experiment $f_\rho^{-1} =$ 0.287(7) and $f_\phi^{-1} =$ 0.234(3).
For comparison use of perturbative $Z_V$ leads to the results 
$f_\rho^{-1} =$ 0.255(27) and $f_\phi^{-1} =$ 0.217(12) which are  
smaller than the experimental values by one standard deviation.

For the axial vector current, a non-perturbative estimate of
$Z_A$ using chiral Ward identities~\cite{MMZMC} 
gives $Z_A=$ 0.741(9), 0.751(11) and 0.822(19) 
for $\beta$=5.9, 6.1 and 6.3~\cite{kuramashi}.  
Using these results and data from ref.~\cite{hashimoto}
we obtain in the continuum limit 
$f_\pi=$ 132(19) MeV and $f_K=$ 154(21) MeV.
Estimating $Z_A$ by 
tadpole-improved perturbation theory, we find 
$f_\pi = $ 122(20) MeV and $f_K = $ 144(18) MeV.
Both methods give results consistent with experiment within errors.

\end{document}